\documentclass[10pt]{article}

\usepackage{lmodern}
\usepackage[T1]{fontenc}
\usepackage[margin=0.78in]{geometry}
\usepackage{amsmath,amssymb}
\usepackage{booktabs}
\usepackage{array}
\usepackage{graphicx}
\usepackage{float}
\usepackage{tikz}
\usepackage{xcolor}
\usepackage{xspace}
\usepackage{hyperref}
\usetikzlibrary{arrows.meta,fit,positioning,shapes.geometric}

\hypersetup{
  colorlinks=true,
  linkcolor=blue!55!black,
  citecolor=blue!55!black,
  urlcolor=blue!55!black
}

\newcommand{\tmem}{\textsc{TMEM}\xspace}
\newcommand{\smem}{\textsc{SMEM}\xspace}

\newcommand{\code}[1]{\texttt{#1}}

\title{Efficient FlashAttention on Blackwell via Fixed-Shift Softmax and
Persistent Scheduling}
\author{Oleksandr Stashuk\\Meta
\and Hongtao Yu\\Meta
\and Jay Shah\\Colfax Research}
\date{Draft, September 2026}

\begin{document}
\maketitle

\begin{abstract}
On Blackwell, normalization and operand movement can limit attention kernels
whose matrix multiplications are already deeply pipelined. We implement an
FA4-style pipeline in Triton TLX with fixed-shift dense softmax and a saved
inverse denominator for backward. The fixed shift removes recurrent
accumulator corrections; the saved reciprocal moves row normalization from
the quadratic backward loop into linear preprocessing. The guarded causal path uses a
first-block anchor with selective recovery, and fully masked diagonal
contractions are skipped. A packed BF16 exponential approximation and
pipelined $dQ$ publication reduce per-element work and on-chip movement.

On B200 at 1,000 W, using prepared launches and the original FA4 timer,
TLX improves geometric-mean throughput by 7.3 percent across the 24 cases
in FA4's BF16 variable-batch grid through 32,768 tokens at head dimension
128. Dense forward and backward improve by 13.0 and 10.4 percent,
respectively. Causal backward improves by 8.0 percent, while causal forward
is 1.5 percent slower. The TritonBench comparison has geometric-mean gains of 13.4 percent
in forward and 24.4 percent in backward.
\end{abstract}

\section{Introduction}

FlashAttention~\cite{dao2022flashattention} reduced attention's high-bandwidth
memory traffic by tiling attention and avoiding materialization of the score
and probability matrices in HBM. Later FlashAttention kernels sought to keep arithmetic units busy
within each tile: FlashAttention-2 improved work
partitioning~\cite{dao2023flashattention2}, FlashAttention-3 overlapped
matrix multiplication and softmax on Hopper~\cite{shah2024flashattention3},
and FlashAttention-4 developed an asynchronous Blackwell pipeline
~\cite{dao2026flashattention4}. On Blackwell, however, tensor-core throughput
has grown faster than shared-memory bandwidth and exponential throughput.
This increases the relative cost of preparing and normalizing tensor-core
operands.

We start from an FA4-style pipeline and examine three constraints on that
overlap. In forward, a changing row maximum requires correction of the
accumulated output before subsequent tiles can use it. In backward,
intermediates share tensor memory (\tmem), so a producer must wait for the previous
value's readers before overwriting its storage. Causal masking adds a
scheduling problem: query tiles have different costs, but reordering them
also changes K/V locality. We address these constraints together through
TLX's asynchronous tasks, warp and register allocations, barriers, and
on-chip storage views~\cite{guan2026tlx,tlxrepository}. The resulting changes are:

\begin{itemize}
  \item Dense dimension-128 forward holds a fixed row shift and generates BF16 weights
  with a packed exponential approximation, eliminating recurrent
  denominator and output-accumulator correction.
  \item Guarded causal forward fixes each row's shift from its first visible key
  block and recomputes tiles that fail numerical checks. Fully masked
  diagonal work is skipped before issuing QK or PV.
  \item Dense dimension-128 forward work is distributed nearly evenly across
  persistent two-CTA clusters. Causal query tiles run in
  reverse order within L2-sized head sections.
  \item Dense dimension-128 forward saves the FP32 inverse denominator.
  Backward scales $dO$ once with this reciprocal and reconstructs
  unnormalized weights, avoiding log-sum-exp (LSE) loads and per-score normalization.
  \item Backward overlaps $dQ$ publication with subsequent matrix
  multiplications. In the measured dimension-128 configurations,
  dense partials use paired BF16 reductions;
  causal lengths from 1,024 through 16,384 use bounded packed fixed-point
  reductions.
\end{itemize}

\newpage
\section{Attention and Blackwell Resource Limits}
\label{sec:problem}

\subsection{Forward and backward equations}

For a batch of $B$ sequences with $H$ attention heads, let
$Q,K,V\in\mathbb{R}^{N\times d}$ denote the inputs for one sequence and head, where $N$ is
the sequence length and $d$ is the head dimension. The caller supplies the scale $\alpha$,
normally $1/\sqrt d$. Indices $i$ and $j$ denote query and key rows, starting
at zero; $Q_i$, $K_j$, and $V_j$ denote row vectors. We use $P$ for exact
softmax probabilities and $O$ for the corresponding exact output. Dense
forward computes
\begin{equation}
\begin{aligned}
  S &= \alpha QK^{\mathsf T},\\
  P_{ij} &= \frac{\exp(S_{ij})}{\sum_k\exp(S_{ik})},\\
  O &= PV.
\end{aligned}
  \label{eq:forward}
\end{equation}
Causal attention sets $S_{ij}=-\infty$ for $j>i$ before normalization, making
those probabilities zero. The kernel folds the supplied scale into score
multiplication or base-two exponent conversion.

For a scalar loss $\mathcal L$, $dX$ denotes its gradient with respect to $X$;
$dO$ is the supplied output gradient. Backward first forms the row reduction
$D$. Using the standard FlashAttention notation, the equations are
\begin{equation}
\begin{aligned}
  D_i &= \sum_t O_{it}dO_{it},\\
  dP &= dOV^{\mathsf T},\\
  dS &= P\odot(dP-D\mathbf{1}^{\mathsf T}),\\
  dV &= P^{\mathsf T}dO,\\
  dQ &= \alpha dS K,\\
  dK &= \alpha dS^{\mathsf T}Q.
\end{aligned}
  \label{eq:backward}
\end{equation}
Here $\odot$ is elementwise multiplication and $\mathbf 1$ is a vector of
ones, so $D_i$ is subtracted from every entry of row $i$. We follow the
FlashAttention algorithms in using $\widetilde P$ for unnormalized shifted
exponentials, $m$ for the row shift, $\ell$ for their row sum, and $L$ for
log-sum-exp~\cite{dao2023flashattention2,shah2024flashattention3,dao2026flashattention4}.
Kernel formulas below reuse this notation for computed quantities, with
rounding stated explicitly. BF16 denotes bfloat16, with eight exponent bits
and seven fraction bits~\cite{kalamkar2019bfloat16}. The operators
$\operatorname{round}_{\mathrm{BF16}}$ and
$\operatorname{round}_{\mathrm{FP32}}$ denote round-to-nearest-even
conversion to the corresponding format.

Score recomputation gives five matrix multiply--accumulate (MMA)
contractions in the fused backward pass. Their overlap depends on when normalization and Jacobian
computations supply the required intermediates; $dK$ and $dQ$ also
consume opposite orientations of $dS$. We therefore retain the five
contractions and target the normalization, operand formats, and storage
reuse that determine when each can begin.

\subsection{Resource limits after the FlashAttention-4 schedule}
\label{sec:resource-limits}

NVIDIA documents Blackwell's on-chip memory and asynchronous tensor-core
operations in its architecture brief, tuning guide, and PTX ISA
reference~\cite{nvidiablackwellbrief,nvidiablackwell,nvidiaptx}.
For a forward tile with 256 query rows, 128 key rows, and head dimension
128, FA4's model assigns 2,048 cycles to $QK^{\mathsf T}$ and $PV$, 1,536 cycles to
shared memory (\smem) operand traffic, and 2,048 cycles to exponential evaluation
~\cite{dao2026flashattention4}. Because exponential evaluation takes as
long as the tensor work, it must overlap the MMAs to avoid extending the
iteration. FA4 combines this overlap with conditional accumulator
rescaling. We further reduce the exponent work by constructing BF16
operands directly with packed fused multiply-add (FMA) and integer instructions, removing
\code{MUFU.EX2} from the dense score path.

An archived profile shows how little tensor-core idle time remains after
these dense transformations. At 750 W with $B=4$, $H=48$, $N=8{,}192$,
and $d=128$, one TLX launch with 14 counter-replay passes recorded
approximately 2,048 tensor-active cycles out of 2,125 elapsed cycles per
streaming multiprocessor (SM) after normalization by the number of 128-key iterations. Tensor-active
cycles agree with FA4's work accounting. At fixed clock and tensor work,
even eliminating all remaining idle cycles would therefore improve this
kernel's throughput by only about 3.7 percent, or reduce latency by
3.6 percent.

Short dense and causal grids pose a different problem. Setup and drain
are amortized over fewer iterations, so uneven work among resident
clusters can dominate the final wave. We choose the dense cluster count
to distribute output tiles nearly evenly. Causal tiles require an
additional ordering decision because their key-loop lengths differ:
we run longer tiles first within L2-sized head sections, preserving
K/V locality while reducing the completion tail.

In backward, FA4's two-CTA model instead puts shared-memory service
above tensor work: 2,688 cycles versus 2,560 cycles for five MMAs.
Of the shared-memory cost, 512 cycles stage and reread FP32 $dQ$
partials. Publishing these partials in BF16 halves that component to
256 cycles, bringing the estimate to 2,432 cycles per SM: 1,536 for
operands, 256 for the $dS$ store, 384 for distributed shared memory (DSMEM),
and 256 for $dQ$.
At $N=8{,}192$, the earlier traffic model adds 8 cycles for row state
and 16 for $dK/dV$ publication, amortized over 64 query tiles, for
2,456 cycles in total. Thus BF16 staging moves the modeled
shared-memory cost below the tensor-core cost, although the amortized
terms grow at shorter sequences.

With this lower traffic estimate, operand readiness becomes the next
scheduling question. An earlier row-normalized TLX kernel with BF16 $dQ$
publication, before BF16 residual arithmetic in $dS$, recorded
75.3 percent tensor activity, 35.2 percent issue activity, 0.53 eligible
warps per scheduler, and 3.0 percent DRAM activity at $B=4$, $H=48$,
$N=8{,}192$, and $d=128$. These counters motivate examining the
operand dependencies: $dV$ waits for probabilities from $QK^{\mathsf T}$, whereas
$dK$ and $dQ$ also need the $dOV^{\mathsf T}$ result to form $dS$. Moreover,
the intermediates must share 512 \tmem columns. We address this
readiness constraint by scheduling contractions around the points at
which their shared storage becomes reusable.

\section{Algorithmic Changes}
\label{sec:observations}

\subsection{Fixed-shift streaming softmax}

For any row constant $m_i$, exact softmax satisfies
\begin{equation}
  \frac{\exp(S_{ij}-m_i)}{\sum_k\exp(S_{ik}-m_i)}=P_{ij},
  \label{eq:gauge}
\end{equation}
because the common factor $\exp(-m_i)$ cancels. As in FA4, the equations
use natural-log units. The kernel converts scores and shifts to base-two
units to evaluate $\exp(S_{ij}-m_i)=2^{(S_{ij}-m_i)\log_2 e}$.
The shift need not equal the row maximum. FlashDecoding++ uses a
precomputed ``unified max value'' to avoid synchronized rescaling of
partial softmax results during decoding~\cite{hong2023flashdecoding}.
Fixed-shift normalization itself is therefore prior work. We apply it
to training forward and backward, combining the packed BF16 evaluator
with reuse of the inverse denominator in backward.
In the standard online recurrence, let $\mathcal{J}_t$ be the key block processed at step $t$.
The running maximum $m_i^{(t)}$, denominator $\ell_i^{(t)}$, and unnormalized
output accumulator $O_i^{(t)}$ are updated as
\begin{equation}
\begin{aligned}
  m_i^{(t)}
    &=\max\!\left(m_i^{(t-1)},\max_{j\in\mathcal{J}_t}S_{ij}\right),\\
  \ell_i^{(t)}
    &=e^{m_i^{(t-1)}-m_i^{(t)}}\ell_i^{(t-1)}
      +\sum_{j\in\mathcal{J}_t}e^{S_{ij}-m_i^{(t)}},\\
  O_i^{(t)}
    &=e^{m_i^{(t-1)}-m_i^{(t)}}O_i^{(t-1)}
      +\sum_{j\in\mathcal{J}_t}e^{S_{ij}-m_i^{(t)}}V_j.
\end{aligned}
  \label{eq:online-recurrence}
\end{equation}
Starting from $m_i^{(-1)}=-\infty$, $\ell_i^{(-1)}=0$, and
$O_i^{(-1)}=0$, the first block supplies at least one finite score.
After the last key block, division by the accumulated row sum gives
the normalized output $O_i$. Whenever the running maximum increases,
however, both accumulators must be rescaled before the next update.
We remove this recurrent dependency in dense BF16 dimension-128
attention by holding a compile-time shift $m$ throughout the row.
Every rescaling factor then equals one, so the correction and the
synchronization that publishes its scale disappear.

This choice trades dynamic range adaptation for a simpler recurrence.
Transformed exponent fields must stay within the evaluator's finite
range, and the row sum must remain finite and nonzero. Negligible tail
terms may underflow within the stated error tolerance, but exponent-field
wraparound is invalid. For guarded causal attention we instead derive
a shift from each row and retain a recovery path, as described next;
Table~\ref{tab:routes} records the measured dispatch choices.

\subsection{Guarded causal normalization}
\label{sec:causal-gauge}

To obtain a row-dependent shift without updating it on every key block,
the guarded route anchors each causal row to the maximum in its first
visible block.
Later blocks may contain larger scores, so this route also checks the
normalization state and can recompute a tile. Let $\mathcal{J}_0(i)$
contain the valid keys in the first block. Showing BF16 weight conversion
while omitting FP32 exponential and accumulation roundoff gives:
\begin{equation}
\begin{aligned}
  m_i &= \max_{j\in\mathcal{J}_0(i)}S_{ij},\\
  \widetilde P_{ij} &= \operatorname{round}_{\mathrm{BF16}}\!\left(e^{S_{ij}-m_i}\right),\\
  \ell_i &= \sum_{j\leq i}e^{S_{ij}-m_i},\\
  O_i &\approx \ell_i^{-1}\sum_{j\leq i}\widetilde P_{ij}V_j.
\end{aligned}
  \label{eq:causal-gauge}
\end{equation}
Native exponentials feed the FP32 denominator before BF16 conversion,
whereas the tensor cores multiply rounded weights by $V$. Consequently,
the approximation sign includes the different operand precision as well
as accumulation and final output rounding. We keep native exponential
evaluation here and reserve the packed evaluator below for the dense path.

Each 128-row group checks its anchor, partial row
sums, final denominator, and accumulated numerator. If either group
fails, a second launch recomputes their entire 256-row query tile with
the conditionally rescaled online path~\cite{hong2023flashdecoding};
otherwise the tile retains the fast result. Both paths save
$L_i=m_i+\log\ell_i$ for backward, using the final online shift and
denominator after recovery. We store this value in base-two units as
$m_i\log_2 e+\log_2\ell_i$, so backward can consume the same state
format from either path. Inputs outside the admitted shapes use online
normalization.

\subsection{Packed BF16 exponential evaluation}
\label{sec:exp}

Once the dense shift $m$ is fixed, we can combine it with score scaling
in the exponent encoding. In base-two units, the integer part of
$m\log_2 e$ scales all weights by a common power of two and provides
range headroom. Although this common scale cancels under exact
normalization, the fractional part changes where scores fall within
the evaluator's approximation and rounding intervals. Forward and
backward therefore use the same shift and reconstruction rule.

PV consumes BF16 operands, so we construct approximate BF16 weights
directly from two FP32 score elements that feed a packed register.
We follow Schraudolph's exponent-field approach~\cite{schraudolph1999}
and related BF16 constructions~\cite{wang2025vexp}; approximate
binary-logarithm arithmetic also has a long history~\cite{mitchell1962logarithms}. Because BF16 has
seven fraction bits, the exponent-coordinate scale is $2^7$. An FP32
fused multiply-add combines this scale, the caller's score scale, and
the fixed shift with an encoding bias. Its rounding component,
$1.5\cdot2^{23}$, exploits unit FP32 spacing to place the rounded
coordinate in the low bits of the result~\cite{muller2016elementary}. A byte
permutation then packs one 16-bit field from each score element into a BF16x2
register. After a packed integer add adjusts the exponent field, one
packed BF16 fused multiply-add supplies the linear mantissa correction
and ReLU clamp.
The corresponding PTX operations include \code{fma.rn.f32x2},
\code{prmt}, and \code{fma.rn.relu.bf16x2}~\cite{nvidiaptx}.

FA4 uses conventional Cody--Waite range reduction before polynomial
evaluation~\cite{dao2026flashattention4,cody1980elementary} and
distributes exponential work between MUFU and that evaluator,
using the latter for 10--25 percent of a row's entries. It evaluates
degrees three through five and finds degree three sufficient after BF16
rounding~\cite{dao2026flashattention4}. We instead use the packed
evaluator for every entry in the dense score loop, leaving final
normalization to the epilogue. Pre-scaling $dO$ lets dense backward
consume the same unnormalized weights;
Section~\ref{sec:rownorm} derives the transformation that removes
score-wise normalization and native exponential evaluation from that loop.

This balance is hardware-dependent. B300 and GB300 provide 32 native
exponential operations per clock per SM~\cite{dao2026flashattention4},
which halves FA4's modeled exponential term to 1,024 cycles when the
other terms are held fixed. Exponential work then falls below the
2,048-cycle MMA term, so its replacement has less modeled headroom to
exploit. Fixed-shift normalization still removes the separate accumulator-correction
dependency.

\subsection{Reducing causal diagonal work}

Skipping fully masked blocks is standard in causal
attention~\cite{dao2023flashattention2}. Here the skip operates within
the boundary tile, omitting its fully masked half.
At the causal boundary, the tile geometry identifies work whose result
is known before any score is computed. Consider a one-CTA work tile
beginning at query $a$, split into two 128-row groups. Queries in the
first group end at $a+127$, whereas keys in the final 128-column block
start at $a+128$. Every score between these groups is therefore masked
to $-\infty$, and every corresponding weight is zero.

We omit the resulting $128\times128\times d$ QK contraction and its
two PV slices over 64 keys each. Each PV slice multiplies a
$128\times64$ weight fragment by a $64\times d$ value fragment,
contracting over key positions and producing all $d$ output columns.
The split is along the key dimension, not the head/output dimension.
Because no weight from this block is consumed, we
can also omit its score load, mask, exponential, row reduction, and
probability stores while preserving the remaining producer--consumer
sequence.

For one batch/head pair and $N$ divisible by 256, let $T=N/256$ be the number
of query tiles, indexed by $t=0,\ldots,T-1$. Tile $t$ reaches key position
$256(t+1)-1$, so it visits $256(t+1)/128=2(t+1)$ key blocks. Each visited
block issues two QK calls, one per query group. Each group also issues two
PV calls with 64-key reduction slices, giving four PV calls. Before diagonal
skipping, the counts are
\begin{equation}
\begin{aligned}
  \text{QK calls} &= \sum_{t=0}^{T-1}2\,[2(t+1)]=2T(T+1),\\
  \text{PV calls} &= \sum_{t=0}^{T-1}4\,[2(t+1)]=4T(T+1).
\end{aligned}
  \label{eq:diagonal-count}
\end{equation}
Each skipped half removes one QK call and two PV calls, or $3T$ of
the $6T(T+1)$ logical contractions. Thus the removed fraction,
$1/[2(T+1)]$, shrinks as sequence length grows. We enable the skip
through $N=4{,}096$ and also at $N=8{,}192$ on the guarded route.

\subsection{Saving the inverse denominator for backward}
\label{sec:rownorm}

Backward normally reconstructs normalized probabilities at every score.
We move their row-wise scale into $dO$ so that one preprocessing pass
can supply it to several contractions. To see why this is possible,
consider exact arithmetic with $\widetilde P_{ij}=e^{S_{ij}-m}$,
$\ell_i=\sum_j\widetilde P_{ij}$, and a finite, positive denominator:
\begin{equation}
\begin{aligned}
  dS_{ij}
    &=P_{ij}(dP_{ij}-D_i)
      =\widetilde P_{ij}\bigl(\ell_i^{-1}dP_{ij}-\ell_i^{-1}D_i\bigr),\\
  dV_j&=\sum_i\widetilde P_{ij}\bigl(\ell_i^{-1}dO_i\bigr).
\end{aligned}
  \label{eq:normalization-factorization}
\end{equation}
Both $dP_{ij}=\sum_t dO_{it}V_{jt}$ and $D_i=\sum_t O_{it}dO_{it}$
are linear in $dO_i$. Scaling $dO_i$ once therefore supplies the reciprocal
to both terms and to $dV$, without applying it to each reconstructed
weight. In the dense implementation, the packed BF16 evaluator and
rounded row reduction give
\begin{equation}
\begin{aligned}
  \widetilde{P}_{ij} &\approx \exp(S_{ij}-m),\\
  \ell_i &\approx\sum_j\widetilde{P}_{ij}.
\end{aligned}
  \label{eq:represented-probability}
\end{equation}
We reduce each 32-value fragment in BF16 and accumulate the totals in FP32,
so the denominator uses weights already rounded to the PV operand format.
Forward then computes $\ell_i^{-1}$ in FP32, uses it to normalize the
output, and saves it for backward. Rounding in both the sum and the
reciprocal means that the represented weights
$\ell_i^{-1}\widetilde P_{ij}$ may sum to more or less than one.

Since backward already reads $dO$ in an $O(Nd)$ preprocessing pass,
we apply the saved reciprocal there before forming $D$. In the following
assignments, $O$ is the saved output; $dO$, $D$, and subsequently $dP$
denote local rescaled operands. We write the scaled gradient to a
separate BF16 buffer:
\begin{equation}
\begin{aligned}
  dO_i &\leftarrow \operatorname{round}_{\mathrm{BF16}}\!\left(
    \operatorname{round}_{\mathrm{FP32}}\!\left(\ell_i^{-1}dO_i\right)\right),\\
  D_i &\leftarrow \sum_t O_{it}dO_{it}.
\end{aligned}
  \label{eq:normalized-preprocess}
\end{equation}
Here the scale product is rounded first to FP32 and then to BF16,
and the second assignment reduces that rounded vector with the saved
output in FP32. Suppressing the remaining BF16 conversions and floating-point
roundoff detailed in Section~\ref{sec:numerics}, the scaled main loop is
\begin{equation}
\begin{aligned}
  dP &\leftarrow dOV^{\mathsf T},\\
  dS &\leftarrow \widetilde{P}\odot(dP-D\mathbf 1^{\mathsf T}),\\
  dV &\leftarrow \widetilde{P}^{\mathsf T}dO,\\
  dQ &\leftarrow \alpha dS K,\\
  dK &\leftarrow \alpha dS^{\mathsf T}Q.
\end{aligned}
  \label{eq:normalized-backward}
\end{equation}
The main loop reconstructs $\widetilde P$ with the forward packed evaluator
and consumes it with the rescaled $dO$, eliminating LSE loads, per-score
LSE subtraction, and native \code{MUFU.EX2} evaluation. Applying the reciprocal
before BF16 conversion changes rounding; Section~\ref{sec:numerics} reports
the resulting gradient error.
With quantized reconstructed weights, the usual softmax Jacobian is
a surrogate for differentiation through the bit-level forward evaluator;
Bengio et al. discuss surrogate-gradient estimators in a different
setting~\cite{bengio2013estimating}.

We likewise fold the score scale into the FP32 QK result. Materializing
a separate scaled BF16 copy of $K$ would add an input-sized pass and
an intermediate rounding step, so we leave $K$ unscaled, including in
the dimension-64 backward route. Corresponding $dQ$ and $dK$ epilogues
apply $\alpha$ consistently with these operands.

\subsection{Bounded query-gradient reduction}
\label{sec:packed-dq}

Since each K/V owner contributes only part of a query gradient, $dQ$
must be reduced across owners. Dense dimension-128 execution publishes
paired BF16 values through a shared-memory ring and combines them with
global reduce-add. For causal lengths from 1,024 through 16,384 at
the same head dimension, we instead pack two signed fixed-point
contributions into one 32-bit integer reduction word. This retains two
components per word while allowing a bound on the added quantization
error and independent recovery of the component sums.

We use the fixed-point scale $F=2{,}048$ through $N=8{,}192$ and
$F=4{,}096$ at $N=16{,}384$. Rounding a scaled FP32
partial to the nearest integer contributes at most $1/(2F)$ of additional
absolute error after conversion back. Each K/V owner covers 256 keys in the
dimension-128 two-CTA route, so query row $i$ receives
$\lfloor i/256\rfloor+1$ contributions. Relative to the exact sum of the
already-computed, $\alpha$-scaled FP32 partials before packing, the
componentwise bound is
\begin{equation}
  \text{fixed-point quantization error in row }i
  \;\leq\;\frac{\lfloor i/256\rfloor+1}{2F}.
  \label{eq:packed-dq}
\end{equation}
For $N=1{,}024$, 2,048, 4,096, 8,192, and 16,384, the last row has 4, 8,
16, 32, and 64 contributors. Equation~\ref{eq:packed-dq} gives bounds of
$9.77\mathrm{e}{-4}$, $1.95\mathrm{e}{-3}$, $3.91\mathrm{e}{-3}$,
$7.81\mathrm{e}{-3}$, and $7.81\mathrm{e}{-3}$, respectively.
At $N=32{,}768$ we use FP32 reduction instead.

The bound follows from the triangle inequality, provided conversion and
each recovered signed 16-bit sum remain within range. We enforce this
condition with runtime amplitude budgets and use carry-correct unpacking
for the packed 32-bit addition. Because the final result is BF16, the
unpacking pass includes that final rounding in its error budget as well.
For the first 256 query rows, only one owner contributes, so packing
serves no reduction purpose: these rows bypass quantization and retain
ordinary direct BF16 rounding. If a range or rounding check fails, we
recompute backward with FP64 $dQ$ accumulation.

\section{TLX Dataflow and Scheduling}
\label{sec:dataflow}

\subsection{On-chip storage}

Blackwell separates asynchronous operand movement, MMA issue, and
accumulator access, but overlapping these operations
requires storage for their live intermediates. At head dimension 128,
we assign each CTA-level forward tile 256 query rows in two 128-row
softmax groups and advance through 128 keys per iteration. The tensor memory
accelerator (TMA) moves
Q/K/V and output tiles, while \tmem holds FP32 scores and output
accumulators; PV consumes weight fragments converted to BF16.
These configurations use about 225 KiB of dynamic \smem and all
512 \tmem columns, leaving room for one CTA per SM.

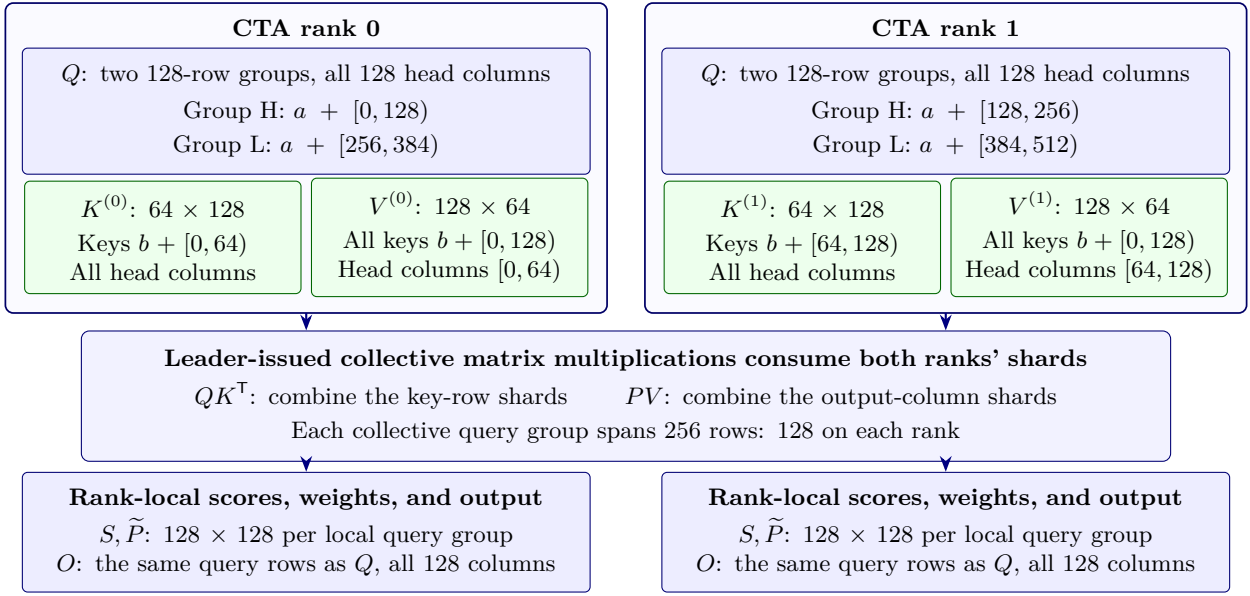
\begin{figure}[htbp]
\centering
\begin{tikzpicture}[
  x=1cm,y=1cm,
  font=\small,
  panel/.style={draw=blue!40!black,fill=blue!2,rounded corners=3pt,line width=0.65pt},
  query/.style={draw=blue!45!black,fill=blue!7,rounded corners=2pt,align=center,inner sep=6pt},
  operand/.style={draw=green!40!black,fill=green!7,rounded corners=2pt,align=center,inner sep=5pt},
  collective/.style={draw=blue!45!black,fill=blue!5,rounded corners=3pt,align=center,inner sep=7pt},
  flow/.style={-{Stealth[length=2.1mm]},draw=blue!50!black,line width=0.8pt}
]
\node[font=\small\bfseries] at (0,0.42)
  {Dense non-causal BF16, $d=128$: one two-CTA cluster owns 512 query rows};
\node[align=center] at (0,-0.02)
  {$a$: first query row \qquad $b$: first key row of the current 128-key block};
\draw[panel] (-8.20,-0.45) rectangle (-0.25,-4.55);
\draw[panel] (0.25,-0.45) rectangle (8.20,-4.55);
\node[font=\small\bfseries] at (-4.225,-0.78) {CTA rank 0};
\node[font=\small\bfseries] at (4.225,-0.78) {CTA rank 1};
\node[query,text width=7.08cm] at (-4.225,-1.88) {
  $Q$: two 128-row groups, all 128 head columns\\[3pt]
  Group H: $a+[0,128)$\\[2pt]
  Group L: $a+[256,384)$
};
\node[query,text width=7.08cm] at (4.225,-1.88) {
  $Q$: two 128-row groups, all 128 head columns\\[3pt]
  Group H: $a+[128,256)$\\[2pt]
  Group L: $a+[384,512)$
};
\node[operand,text width=3.30cm,minimum height=1.38cm] at (-6.12,-3.55) {
  $K^{(0)}$: $64\times128$\\[2pt]
  Keys $b+[0,64)$\\
  All head columns
};
\node[operand,text width=3.30cm,minimum height=1.38cm] at (-2.33,-3.55) {
  $V^{(0)}$: $128\times64$\\[2pt]
  All keys $b+[0,128)$\\
  Head columns $[0,64)$
};
\node[operand,text width=3.30cm,minimum height=1.38cm] at (2.33,-3.55) {
  $K^{(1)}$: $64\times128$\\[2pt]
  Keys $b+[64,128)$\\
  All head columns
};
\node[operand,text width=3.30cm,minimum height=1.38cm] at (6.12,-3.55) {
  $V^{(1)}$: $128\times64$\\[2pt]
  All keys $b+[0,128)$\\
  Head columns $[64,128)$
};
\node[collective,text width=13.90cm,minimum height=1.22cm] (fc-mma) at (0,-5.65) {
  \textbf{Leader-issued collective matrix multiplications consume both ranks' shards}\\[3pt]
  $QK^{\mathsf T}$: combine the key-row shards \qquad
  $PV$: combine the output-column shards\\[2pt]
  Each collective query group spans 256 rows: 128 on each rank
};
\coordinate (fc-in0) at (-4.225,-4.55);
\coordinate (fc-in1) at (4.225,-4.55);
\draw[flow] (fc-in0) -- (fc-mma.north -| fc-in0);
\draw[flow] (fc-in1) -- (fc-mma.north -| fc-in1);
\node[query,text width=7.08cm,minimum height=1.18cm] (fc-out0) at (-4.225,-7.45) {
  \textbf{Rank-local scores, weights, and output}\\[3pt]
  $S,\widetilde P$: $128\times128$ per local query group\\
  $O$: the same query rows as $Q$, all 128 columns
};
\node[query,text width=7.08cm,minimum height=1.18cm] (fc-out1) at (4.225,-7.45) {
  \textbf{Rank-local scores, weights, and output}\\[3pt]
  $S,\widetilde P$: $128\times128$ per local query group\\
  $O$: the same query rows as $Q$, all 128 columns
};
\draw[flow] (fc-mma.south -| fc-out0.north) -- (fc-out0.north);
\draw[flow] (fc-mma.south -| fc-out1.north) -- (fc-out1.north);
\end{tikzpicture}
\caption{Ownership in the dense non-causal BF16 two-CTA forward configuration. A CTA is a cooperative
thread array, or CUDA thread block. Each CTA owns 256 query rows in two
interleaved 128-row groups. Following FA4, H and L denote the first (upper)
and second (lower) query groups. Corresponding groups across
the two ranks form a collective
contraction. The CTA pair partitions K/V into the shown shards and reuses
them across both query groups. Softmax and row reductions remain local
to each query row.}
\label{fig:forward-two-cta-ownership}
\end{figure}

For the dense non-causal BF16 route at $d=128$, we pair CTAs into a cluster covering 512
query rows. Partitioning K/V across the pair lets collective matrix
multiplications consume both ranks' shards while keeping scores and
output rows local, as shown in Figure~\ref{fig:forward-two-cta-ownership}.
Causal forward uses one CTA per work tile, including the guarded and
recovery paths. Forward at head dimension 64 also uses one CTA. To support both head-major
$[B,H,N,d]$ and sequence-major $[B,N,H,d]$ inputs, we address their
strides directly through tensor descriptors and preserve the layout
in outputs and gradients.

Because $dK$ and $dV$ accumulate across query tiles, we reserve two
128-column \tmem regions for them in backward. The remaining 256
columns must serve the transient intermediates, so we divide them into
two regions and reuse each according to operand lifetime. Transposed
scores, $\widetilde P^{\mathsf T}$, and the $dQ$ partial share one;
$dP^{\mathsf T}$ and $dS^{\mathsf T}$ share the other. Replacing a
value requires all of its readers to finish. We enforce these lifetime
boundaries with named barriers between asynchronous tasks.

\subsection{Measured configurations}
\label{sec:configurations}

Table~\ref{tab:routes} records which normalization and reduction paths
the benchmarked binaries use. Only dense dimension-128 forward saves
the reciprocal denominator; the other paths save LSE for reconstruction.
Work acquisition also differs by route: dense dimension-128 forward
uses a fixed-stride persistent grid, other main forward kernels use
cluster launch control, and sparse recovery maps directly to flagged
tiles. In backward, the packed causal route is persistent and the
remaining routes use nonpersistent grids.

\begin{table}[htbp]
\centering
\caption{Measured BF16 kernel paths. CTA counts are per cluster.
Guarded forward uses the first-block shift from
Section~\ref{sec:causal-gauge}, followed by sparse online recovery.
Here \emph{all} denotes the six tested lengths in Section~\ref{sec:fa4-config}.}
\label{tab:routes}
\small
\setlength{\tabcolsep}{4pt}
\begin{tabular}{rlllcl}
\toprule
$d$ & mask & $N$ & Forward normalization & CTAs fwd/bwd & $dQ$ reduction \\
\midrule
128 & dense & all & fixed, packed exponential & 2/2 & BF16 \\
128 & causal & 1,024 & online, native exponential & 1/2 & packed integer \\
128 & causal & 2,048--16,384 & guarded, native exponential & 1/2 & packed integer \\
128 & causal & 32,768 & guarded, native exponential & 1/2 & FP32 \\
64 & dense & all & online, native exponential & 1/1 & FP32 \\
64 & causal & 1,024 & online, native exponential & 1/1 & FP32 \\
64 & causal & 2,048--32,768 & guarded, native exponential & 1/1 & FP32 \\
\bottomrule
\end{tabular}
\end{table}

Dense dimension-64 forward uses conditional rescaling at 2,048, 4,096,
and 16,384 tokens and ordinary online rescaling at the other lengths.
Its backward path uses a separate layout postprocessing pass.
Dense dimension-128 backward uses two $dQ$ staging buffers at
$N=8{,}192$ and one at the other lengths.

Table~\ref{tab:warp-budgets} gives the task decomposition. Register
requests are source-level controls per thread. The compiler manages
the default task's budget and allocates the physical warp slots.

\begin{table}[htbp]
\centering
\caption{Warp roles and register requests in the measured main kernels.
Each additional compute-task entry gives warps and requested registers
per thread; one-warp roles share the stated request unless noted.
The final column gives physical warps per CTA from the loaded binaries.}
\label{tab:warp-budgets}
\footnotesize
\setlength{\tabcolsep}{4pt}
\begin{tabular}{>{\raggedright\arraybackslash}p{2.3cm}>{\raggedright\arraybackslash}p{3.1cm}>{\raggedright\arraybackslash}p{3.35cm}>{\raggedright\arraybackslash}p{4.15cm}r}
\toprule
Kernel & Default task & Additional compute tasks & One-warp roles & Warps \\
\midrule
Dense $d=128$ forward &
4 warps, first softmax group &
Second softmax group: 4 warps, 176 registers &
MMA, first-K load, Q/K/V load, output: 24 registers each &
12 \\
Guarded causal forward &
4 warps, first softmax group &
Second softmax group: 4 warps, 232 registers &
MMA, Q/K/V load, output: 24 registers each &
12 \\
Online / recovery forward &
4 warps, correction and normalization &
Two softmax groups: 4 warps each; 168 registers online, 176 in recovery &
MMA, Q/K/V load, output: 24 registers each &
16 \\
Direct $d=128$ backward &
8 warps, weights, $dS$, exchange, $dK/dV$ epilogue &
$dQ$ staging: 4 warps, 88 registers &
MMA, operand load, $dQ$ output: 88 each; DSMEM relay: 40 &
16 \\
$d=64$ backward &
8 warps, weights, $dS$, $dK/dV$ epilogue &
$dQ$ publication: 4 warps, 88 registers &
MMA, operand load: 88 registers each &
16 \\
\bottomrule
\end{tabular}
\end{table}

\subsection{Forward pipeline}

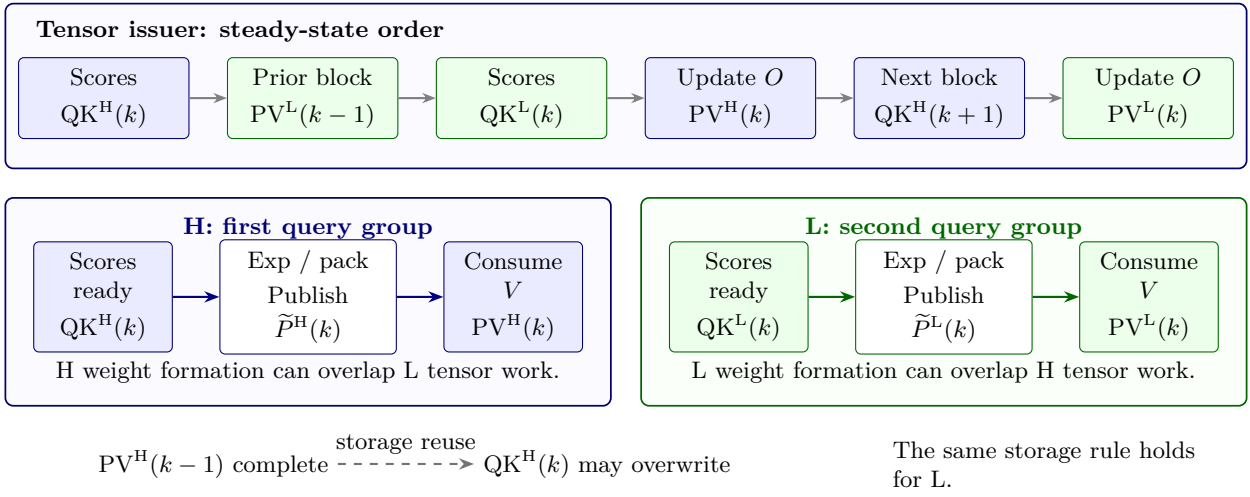
\begin{figure}[t]
\centering
\begin{tikzpicture}[
  x=1cm,y=1cm,
  font=\small,
  panel/.style={draw=blue!40!black,fill=blue!2,rounded corners=3pt,line width=0.65pt},
  tensor/.style={rounded corners=2pt,align=center,text width=1.97cm,minimum height=1.04cm,inner sep=4pt},
  hgroup/.style={draw=blue!45!black,fill=blue!8},
  lgroup/.style={draw=green!40!black,fill=green!8},
  fragment/.style={rounded corners=2pt,align=center,minimum height=1.14cm,inner sep=4pt},
  hflow/.style={-{Stealth[length=2mm]},draw=blue!50!black,line width=0.8pt},
  lflow/.style={-{Stealth[length=2mm]},draw=green!40!black,line width=0.8pt},
  order/.style={-{Stealth[length=1.8mm]},draw=black!50,line width=0.8pt},
  reuse/.style={-{Stealth[length=2mm]},draw=black!55,dashed,line width=0.8pt}
]
\draw[panel] (-8.20,3.30) rectangle (8.20,1.12);
\node[anchor=west,font=\small\bfseries] at (-7.92,2.94)
  {Tensor issuer: steady-state order};
\node[tensor,hgroup] (fo-qh) at (-6.90,2.05)
  {Scores\\[3pt]$\mathrm{QK}^{\mathrm H}(k)$};
\node[tensor,lgroup] (fo-plold) at (-4.14,2.05)
  {Prior block\\[3pt]$\mathrm{PV}^{\mathrm L}(k-1)$};
\node[tensor,lgroup] (fo-ql) at (-1.38,2.05)
  {Scores\\[3pt]$\mathrm{QK}^{\mathrm L}(k)$};
\node[tensor,hgroup] (fo-ph) at (1.38,2.05)
  {Update $O$\\[3pt]$\mathrm{PV}^{\mathrm H}(k)$};
\node[tensor,hgroup] (fo-qhnext) at (4.14,2.05)
  {Next block\\[3pt]$\mathrm{QK}^{\mathrm H}(k+1)$};
\node[tensor,lgroup] (fo-pl) at (6.90,2.05)
  {Update $O$\\[3pt]$\mathrm{PV}^{\mathrm L}(k)$};
\draw[order] (fo-qh) -- (fo-plold);
\draw[order] (fo-plold) -- (fo-ql);
\draw[order] (fo-ql) -- (fo-ph);
\draw[order] (fo-ph) -- (fo-qhnext);
\draw[order] (fo-qhnext) -- (fo-pl);

\draw[panel] (-8.20,0.73) rectangle (-0.20,-2.02);
\draw[panel,draw=green!40!black,fill=green!2] (0.20,0.73) rectangle (8.20,-2.02);
\node[font=\small\bfseries,text=blue!45!black] at (-4.20,0.33)
  {H: first query group};
\node[font=\small\bfseries,text=green!40!black] at (4.20,0.33)
  {L: second query group};
\node[fragment,hgroup,text width=1.55cm] (fo-hscore) at (-6.91,-0.58)
  {Scores ready\\[3pt]$\mathrm{QK}^{\mathrm H}(k)$};
\node[fragment,draw=blue!45!black,fill=white,text width=2.05cm] (fo-hprob) at (-4.20,-0.58)
  {Exp / pack\\[2pt]Publish\\[2pt]$\widetilde P^{\mathrm H}(k)$};
\node[fragment,hgroup,text width=1.55cm] (fo-hpv) at (-1.49,-0.58)
  {Consume $V$\\[3pt]$\mathrm{PV}^{\mathrm H}(k)$};
\node[fragment,lgroup,text width=1.55cm] (fo-lscore) at (1.49,-0.58)
  {Scores ready\\[3pt]$\mathrm{QK}^{\mathrm L}(k)$};
\node[fragment,draw=green!40!black,fill=white,text width=2.05cm] (fo-lprob) at (4.20,-0.58)
  {Exp / pack\\[2pt]Publish\\[2pt]$\widetilde P^{\mathrm L}(k)$};
\node[fragment,lgroup,text width=1.55cm] (fo-lpv) at (6.91,-0.58)
  {Consume $V$\\[3pt]$\mathrm{PV}^{\mathrm L}(k)$};
\draw[hflow] (fo-hscore) -- (fo-hprob);
\draw[hflow] (fo-hprob) -- (fo-hpv);
\draw[lflow] (fo-lscore) -- (fo-lprob);
\draw[lflow] (fo-lprob) -- (fo-lpv);
\node[font=\small] at (-4.20,-1.57)
  {H weight formation can overlap L tensor work.};
\node[font=\small] at (4.20,-1.57)
  {L weight formation can overlap H tensor work.};

\node[align=center] (fo-old-use) at (-5.45,-2.79)
  {$\mathrm{PV}^{\mathrm H}(k-1)$ complete};
\node[align=center] (fo-new-write) at (-0.25,-2.79)
  {$\mathrm{QK}^{\mathrm H}(k)$ may overwrite};
\draw[reuse] (fo-old-use) -- node[above,font=\small] {storage reuse} (fo-new-write);
\node[align=left,text width=4.35cm,anchor=west] at (3.40,-2.79)
  {The same storage rule holds for L.};
\end{tikzpicture}
\caption{Forward pipeline for the dense non-causal H/L query groups defined in
Figure~\ref{fig:forward-two-cta-ownership}. The top row retains FA4's MMA issue
order~\cite{dao2026flashattention4}, with each PV box grouping
32/32/64-key slices along the contracting dimension. These slices are
distinct from each rank's 64 output-head-column shard of $V$. Within each group
(middle panels), publishing the first weight fragment allows the first
PV slice to begin; later slices wait for their own fragments and V.
The groups synchronize independently, while two-CTA execution requires
both ranks to publish each fragment before its PV slice proceeds.
The dashed edge marks when aliased storage becomes reusable.}
\label{fig:forward-overlap-dependencies}
\end{figure}

We retain FA4's forward MMA issue order, interleaving the two query
groups, H and L~\cite{dao2026flashattention4}. After issuing H's QK for key block $k$, the
issuer can run L's PV from block $k-1$ while H forms its weights.
It then issues L's QK followed by H's PV, giving L the corresponding
opportunity to overlap softmax with H's tensor work.
Figure~\ref{fig:forward-overlap-dependencies} shows this steady-state
order and the dependencies within each group. Here QK and PV name
contraction roles; fixed-shift PV multiplies $\widetilde P$ by $V$.
TMA producers, softmax workers, the MMA issuer, accumulator control,
and output publication are assigned to asynchronous tasks with the
warp assignments in Table~\ref{tab:warp-budgets}.

Storage reuse constrains how far those tasks can advance. A probability
fragment can overwrite its source scores only after their final read,
and PV must wait for that fragment to be published. Fixing the shift
removes the additional rescaling factor, its publication synchronization,
and the FP32 accumulator correction. Nevertheless, the
QK--softmax--PV dependency remains exposed on short causal tiles,
where less independent work is available to overlap it. The register
and on-chip storage footprint also limits resident CTAs.

At the end of a query tile, the epilogue multiplies each FP32 output
row by $\ell_i^{-1}$, converts it to the output type, stages it in
\smem, and issues TMA stores. We release output publication before
storing the independent FP32 reciprocal so that this store can overlap
the following work.

To amortize task setup over several output tiles,
each dense two-CTA cluster advances by a fixed grid stride. We bound
the cluster count on the host to distribute those tiles nearly evenly,
reducing final-wave imbalance.

\subsection{Causal scheduling}

Causal query tiles have increasing mainloop lengths, so processing them
left to right leaves the longest work for last. Longest-processing-time-first
(LPT) scheduling~\cite{graham1969bounds} is a heuristic for reducing this tail, but a global
ordering can separate consumers of the same K/V region. We therefore
follow FA4's L2-constrained LPT schedule~\cite{dao2026flashattention4}:
size head sections from the K/V working-set estimate, then process query
blocks in reverse order within each section. Because each head has its
own K/V tensors, the section must be small enough for query tiles to
reuse their head's cached operands. Its size consequently depends on
head dimension, sequence length, and the number of heads sharing the
cache budget.

For guarded causal forward, we encode this section-local ordering directly in the launch grid.
One coordinate exposes a head index, while the second encodes the
remaining query, head-section, and batch coordinates. This reduces
index-decoding work compared with a fully flattened batch--head--query
index.

\subsection{Backward contractions and accumulator flow}

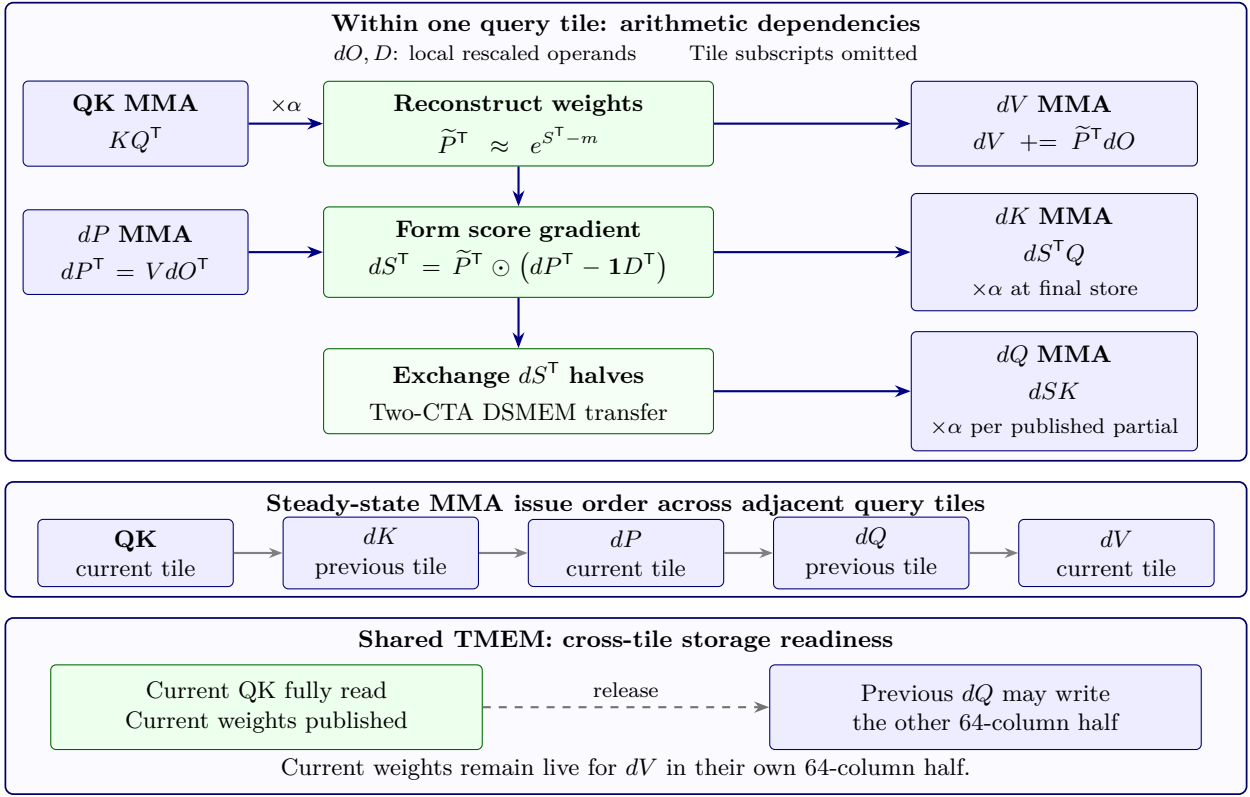
\begin{figure}[H]
\centering
\begin{tikzpicture}[
  x=1cm,y=1cm,
  font=\small,
  panel/.style={draw=blue!40!black,fill=blue!2,rounded corners=3pt,line width=0.65pt},
  mma/.style={draw=blue!45!black,fill=blue!7,rounded corners=2pt,align=center,inner sep=5pt,minimum height=1.12cm},
  scalar/.style={draw=green!40!black,fill=green!7,rounded corners=2pt,align=center,inner sep=5pt,minimum height=1.12cm},
  exchange/.style={draw=green!40!black,fill=green!4,rounded corners=2pt,align=center,inner sep=5pt,minimum height=1.12cm},
  issue/.style={draw=blue!45!black,fill=blue!7,rounded corners=2pt,align=center,inner sep=4pt,minimum height=0.72cm,text width=2.30cm},
  flow/.style={-{Stealth[length=2.1mm]},draw=blue!50!black,line width=0.8pt},
  order/.style={-{Stealth[length=1.8mm]},draw=black!50,line width=0.8pt},
  reuse/.style={-{Stealth[length=2mm]},draw=black!55,dashed,line width=0.8pt}
]
\draw[panel] (-8.20,0.12) rectangle (8.20,-5.94);
\node[font=\small\bfseries] at (0,-0.18)
  {Within one query tile: arithmetic dependencies};
\node[font=\footnotesize] at (0,-0.57)
  {$dO,D$: local rescaled operands \qquad Tile subscripts omitted};
\node[mma,text width=2.62cm] (bp-qk) at (-6.48,-1.48) {
  \textbf{QK MMA}\\[3pt]
  $KQ^{\mathsf T}$
};
\node[scalar,text width=4.80cm] (bp-p) at (-1.42,-1.48) {
  \textbf{Reconstruct weights}\\[3pt]
  $\widetilde P^{\mathsf T}\approx e^{S^{\mathsf T}-m}$
};
\node[mma,text width=3.43cm] (bp-dv) at (5.66,-1.48) {
  \textbf{$dV$ MMA}\\[3pt]
  $dV\mathrel{+}=\widetilde P^{\mathsf T}dO$
};
\node[mma,text width=2.62cm] (bp-dp) at (-6.48,-3.18) {
  \textbf{$dP$ MMA}\\[3pt]
  $dP^{\mathsf T}=VdO^{\mathsf T}$
};
\node[scalar,text width=4.80cm] (bp-ds) at (-1.42,-3.18) {
  \textbf{Form score gradient}\\[3pt]
  $dS^{\mathsf T}=\widetilde P^{\mathsf T}\odot\bigl(dP^{\mathsf T}-\mathbf 1D^{\mathsf T}\bigr)$
};
\node[mma,text width=3.43cm] (bp-dk) at (5.66,-3.18) {
  \textbf{$dK$ MMA}\\[3pt]
  $dS^{\mathsf T}Q$\\[2pt]
  {\footnotesize $\times\alpha$ at final store}
};
\node[exchange,text width=4.80cm] (bp-exchange) at (-1.42,-5.02) {
  \textbf{Exchange $dS^{\mathsf T}$ halves}\\[3pt]
  Two-CTA DSMEM transfer
};
\node[mma,text width=3.43cm] (bp-dq) at (5.66,-5.02) {
  \textbf{$dQ$ MMA}\\[3pt]
  $dS K$\\[2pt]
  {\footnotesize $\times\alpha$ per published partial}
};
\draw[flow] (bp-qk) -- node[above,font=\footnotesize] {$\times\alpha$} (bp-p);
\draw[flow] (bp-p) -- (bp-dv);
\draw[flow] (bp-p) -- (bp-ds);
\draw[flow] (bp-dp) -- (bp-ds);
\draw[flow] (bp-ds) -- (bp-dk);
\draw[flow] (bp-ds) -- (bp-exchange);
\draw[flow] (bp-exchange) -- (bp-dq);
\draw[panel] (-8.20,-6.22) rectangle (8.20,-7.74);
\node[font=\small\bfseries] at (0,-6.52)
  {Steady-state MMA issue order across adjacent query tiles};
\node[issue] (bp-issue-qk) at (-6.48,-7.16) {\textbf{QK}\\current tile};
\node[issue] (bp-issue-dk) at (-3.24,-7.16) {\textbf{$dK$}\\previous tile};
\node[issue] (bp-issue-dp) at (0,-7.16) {\textbf{$dP$}\\current tile};
\node[issue] (bp-issue-dq) at (3.24,-7.16) {\textbf{$dQ$}\\previous tile};
\node[issue] (bp-issue-dv) at (6.48,-7.16) {\textbf{$dV$}\\current tile};
\draw[order] (bp-issue-qk) -- (bp-issue-dk);
\draw[order] (bp-issue-dk) -- (bp-issue-dp);
\draw[order] (bp-issue-dp) -- (bp-issue-dq);
\draw[order] (bp-issue-dq) -- (bp-issue-dv);
\draw[panel] (-8.20,-8.02) rectangle (8.20,-10.35);
\node[font=\small\bfseries] at (0,-8.33)
  {Shared TMEM: cross-tile storage readiness};
\node[scalar,text width=5.35cm] (bp-qk-release) at (-4.75,-9.20)
  {Current QK fully read\\Current weights published};
\node[mma,text width=5.35cm] (bp-dq-reuse) at (4.75,-9.20)
  {Previous $dQ$ may write\\the other 64-column half};
\draw[reuse] (bp-qk-release) -- node[above,font=\footnotesize] {release} (bp-dq-reuse);
\node[font=\small] at (0,-10.02)
  {Current weights remain live for $dV$ in their own 64-column half.};
\end{tikzpicture}
\caption{Backward dependencies and issue order for the dense two-CTA route,
following FA4's issue order and temporary-storage aliasing~\cite{dao2026flashattention4}.
In the upper panel, blue boxes identify the five matrix multiplications; green boxes reconstruct
weights, form the residual, or exchange operands. Only the $dQ$ branch needs
the shown DSMEM exchange. The matrix products for
$dQ$ and $dK$ are unscaled, with $\alpha$ applied once during publication.
The middle strip orders MMA issue across adjacent query tiles.
The dashed edge shows the current QK read-release required before the
previous tile's $dQ$ can reuse half of its storage.}
\label{fig:backward-pipeline}
\end{figure}

The MMA issue order and temporary \tmem aliasing follow FA4's two-CTA
backward pipeline~\cite{dao2026flashattention4}. To accumulate $dK$ and $dV$ across query tiles, a dimension-128 backward
task keeps ownership of two 128-row K/V tiles across a two-CTA cluster
and streams 128-row Q/$dO$ tiles through them. This ownership makes
key rows the leading dimension of the transposed score-like
intermediates. In the dense inverse-denominator route, $dO$, $dP$,
and $D$ are the local rescaled operands from Section~\ref{sec:rownorm}.
Figure~\ref{fig:backward-pipeline} relates their data dependencies to
the five-contraction issue order.

Following FA4, we issue the current tile's QK before the previous tile's $dK$, then
current $dP$, previous $dQ$, and current $dV$. Placing current $dP$
between previous $dK$ and $dQ$ allows its MMA to overlap the DSMEM
exchange required by the previous tile's $dQ$. This interleaving lets
weight and residual formation overlap MMAs whose operands are already
available. QK occupies a 128-column temporary \tmem region. After all
eight compute warps finish reading it and publish the weights, the
current $\widetilde P^{\mathsf T}$ and previous $dQ$ coexist in separate
64-column halves. The $dQ$ issuer waits for this current QK read-release
as well as the previous tile's $dS$ exchange.

The other temporary region holds $dP^{\mathsf T}$, then
$dS^{\mathsf T}$. A named barrier ensures that all eight compute warps
have read $dP^{\mathsf T}$ before it is overwritten. The next $dP$ also
waits until $dK$ has consumed the previous $dS^{\mathsf T}$ and the
compute task has captured its exchange data.

Producing $dQ$ requires an additional exchange. Each CTA owns only
half of the transposed $dS$ tile, but its $dQ$ rows need the full
reduction dimension. The pair therefore exchanges halves through
DSMEM before forming those partials.

FA4 already releases $dQ$'s \tmem storage after copying its FP32
partials into registers, before FP32 global reduce-add. We retain this
early read-release: four reduction warps read the unscaled MMA result
into registers. A named rendezvous marks
that point, allowing later tensor work to reuse the region while the
warps apply $\alpha$, convert the partial, and stage slices in a
shared-memory ring. A separate one-warp task issues TMA stores with
global reduce-add while the next slice is prepared. Thus publication
can overlap subsequent contractions without retaining $dQ$ in \tmem.
Slice width and buffer depth depend on head dimension, reduction format,
and concurrent MMA work; the CTA holds only one or a small ring of
query-tile partials, while the full $N\times d$ tensor stays in global
memory. We likewise apply $\alpha$ to $dK$ after accumulation.

Packed causal backward keeps a resident cluster across multiple K/V
owners, whereas dense backward uses a nonpersistent grid. Both overlap
publication with contraction work within each owner. At dimension 64,
one CTA streams the same 128-row query tiles with two Q buffers and
the same five-contraction topology.

Our changes target work and traffic on these existing barrier paths:
fixed-shift packed reconstruction before $S\to\widetilde P$
publication, BF16 residual arithmetic before $dP\to dS$ publication,
and incremental BF16 $dQ$ publication on the drain path. The required
overwrite and read-release synchronization remains. The five-contraction
order and early \tmem read-release are inherited from FA4.

\section{Numerical Behavior}
\label{sec:numerics}

\subsection{Finite-precision operations}

Forward rounding affects backward through the saved output and normalization
state. Dense
dimension-128 forward constructs packed BF16 exponentials directly.
Backward scales $dO$ by the saved reciprocal and rounds it to BF16
before the $dP$ and $dV$ MMAs; it also rounds $dP-D$ to BF16 and
multiplies that residual by $\widetilde P$ in BF16. Because backward
applies the standard softmax Jacobian to these approximate weights,
reconstruction, reciprocal scaling, and residual rounding introduce
error beyond final BF16 conversion. We measure the resulting gradients
against an independent FP32 attention reference.

Publication and causal execution introduce additional numerical effects.
Direct $dQ$ publication changes the order and precision of global
reduction, while packed causal $dQ$ adds the bounded fixed-point error
in Section~\ref{sec:packed-dq}. Guarded causal forward fixes the shift
after the first block. By contrast, the diagonal skip removes
contractions whose weights are zero under the causal mask; its
regression checks preserve outputs bitwise.

Short causal prefixes need particular care because rounding the saved
output changes the row correction used in backward. Following Qiu and
Yao's analysis~\cite{qiu2026lowprecision}, we recompute
$D_i=\sum_jP_{ij}dP_{ij}$ from reconstructed probabilities and $dP$
for the first 128 query rows. We also add the low-precision
probability-residual contribution to $dV$ at dimension 64 and in the
dimension-128 case at $N=1{,}024$. Limiting these corrections to the
affected prefixes partially compensates rounding without increasing
precision across the full quadratic domain.

\subsection{Intrinsic exponential accuracy}
\label{sec:intrinsic-exp}

We test the four-stage packed evaluator from Section~\ref{sec:exp}
at unit base-two score scale, retaining its encoding bias and correction
constants. Following FA4's interval and sample-count convention, we draw
four million exponents uniformly from the 21-bit FP32 lattice in $[0,1)$,
with seed 20260915. We choose exact FP32 input preimages so that the
unrounded affine encoding represents each sampled exponent. The reference
is FP64 base-two exponential evaluation; the ideal BF16 result uses a
single round-to-nearest-even conversion.

Table~\ref{tab:intrinsic-exp} summarizes the intrinsic exponential errors.

\begin{table}[htbp]
\centering
\caption{Intrinsic relative error against FP64 exponentials on four million
samples. The packed evaluator returns BF16 directly.}
\label{tab:intrinsic-exp}
\small
\begin{tabular}{lrr}
\toprule
Evaluator & Maximum relative error & Mean relative error \\
\midrule
Ideal FP64 to BF16 & $3.89\mathrm{e}{-3}$ & $1.41\mathrm{e}{-3}$ \\
Packed BF16 evaluator & $6.87\mathrm{e}{-2}$ & $5.79\mathrm{e}{-2}$ \\
\bottomrule
\end{tabular}
\end{table}

Packed outputs exceed the FP64 reference by a factor ranging from
1.0440 to 1.0687. Exact row normalization cancels a common multiplicative
factor, but this ratio varies with the exponent, so cancellation leaves
input-dependent error and subsequent rounding effects. Repeating the
test over one period with the deployed nominal shift gives maximum
and mean relative errors of $6.93\mathrm{e}{-2}$ and
$5.85\mathrm{e}{-2}$, again above the BF16 rounding floor.
To determine how these errors affect attention, the checks below
include dot-product evaluation, normalization, and gradient formation.

\subsection{Accuracy of the benchmarked kernels}
\label{sec:current-accuracy}

Because backward consumes state produced by forward, we qualify each
provider using its own saved forward state. These checks use the same
binaries and inputs as the 1,000 W measurements in
Section~\ref{sec:fa4-config}. An independent reference computes chunked
QK, softmax, PV, and analytical backward in FP32 with TF32 disabled.
It forms $D$ from $P\odot dP$, while error metrics accumulate in FP64.
As a consistency check, comparison with direct autograd on a small
case gives a maximum difference of $9.54\mathrm{e}{-7}$.

Both providers pass $|a-b|\leq0.01+0.01|b|$ for output and gradients
at all six dimension-128 lengths through 32,768, with both masks.
We replay backward ten times using each provider's own forward state
and check TLX output, normalizer, and recovery-map repeatability over
ten forward replays. To show the errors within this acceptance
criterion, Table~\ref{tab:current-accuracy} reports maximum absolute
error and relative $\ell_2$ error, $\|a-b\|_2/\|b\|_2$,
against the FP32 reference.

\begin{table}[htbp]
\centering
\caption{Accuracy of the dimension-128 benchmarked binaries against the
FP32 reference. Each entry is the maximum over six lengths and all
recorded checks and replays; different columns may attain their maxima
at different lengths. Acceptance uses the combined absolute/relative
criterion stated in the text.}
\label{tab:current-accuracy}
\small
\begin{tabular}{llrrrr}
\toprule
 & & \multicolumn{2}{c}{Maximum absolute error} &
 \multicolumn{2}{c}{Relative $\ell_2$ error} \\
Mask & Tensor & FA4 & TLX & FA4 & TLX \\
\midrule
Dense & $O$ & $2.90\mathrm{e}{-3}$ & $4.90\mathrm{e}{-3}$ & $2.35\mathrm{e}{-3}$ & $5.21\mathrm{e}{-3}$ \\
Dense & $dQ$ & $3.46\mathrm{e}{-3}$ & $8.86\mathrm{e}{-3}$ & $2.37\mathrm{e}{-3}$ & $1.47\mathrm{e}{-2}$ \\
Dense & $dK$ & $3.55\mathrm{e}{-3}$ & $7.39\mathrm{e}{-3}$ & $2.35\mathrm{e}{-3}$ & $5.94\mathrm{e}{-3}$ \\
Dense & $dV$ & $2.52\mathrm{e}{-3}$ & $5.19\mathrm{e}{-3}$ & $2.35\mathrm{e}{-3}$ & $5.46\mathrm{e}{-3}$ \\
Causal & $O$ & $1.02\mathrm{e}{-2}$ & $1.02\mathrm{e}{-2}$ & $2.17\mathrm{e}{-3}$ & $2.16\mathrm{e}{-3}$ \\
Causal & $dQ$ & $1.69\mathrm{e}{-2}$ & $1.69\mathrm{e}{-2}$ & $2.37\mathrm{e}{-3}$ & $1.27\mathrm{e}{-2}$ \\
Causal & $dK$ & $1.47\mathrm{e}{-2}$ & $1.47\mathrm{e}{-2}$ & $2.36\mathrm{e}{-3}$ & $2.36\mathrm{e}{-3}$ \\
Causal & $dV$ & $2.20\mathrm{e}{-2}$ & $2.20\mathrm{e}{-2}$ & $2.33\mathrm{e}{-3}$ & $2.33\mathrm{e}{-3}$ \\
\bottomrule
\end{tabular}
\end{table}

TLX's largest dense $dQ$ relative $\ell_2$ error occurs at 32,768 tokens:
$1.47\mathrm{e}{-2}$ for TLX and $2.37\mathrm{e}{-3}$ for FA4.
At that length, TLX's dense maximum absolute errors in
$(O,dQ,dK,dV)$ are
$(3.59\mathrm{e}{-4},3.43\mathrm{e}{-3},7.08\mathrm{e}{-4},4.10\mathrm{e}{-4})$.
The causal values are
$(7.86\mathrm{e}{-3},1.12\mathrm{e}{-2},1.14\mathrm{e}{-2},1.69\mathrm{e}{-2})$.
Thus the faster implementation has a measurable precision cost,
especially in $dQ$.

\subsection{Recovery observations}

Recovery flags describe 128-row query groups, while recomputation covers
a 256-row tile whenever either of its two groups is flagged.
Every guarded dimension-128 case records zero flagged groups over
the ten qualification replays and post-timing checks. A separate
dimension-64 check with $B=4$ and $H=48$ does trigger recovery:
the 16,384-token case flags one of 24,576 groups,
and the 32,768-token case flags five of 49,152. Each recorded map
repeats across the probes. The timed callback includes the recovery launch
even when every flag is zero.

\subsection{Limitations}
\label{sec:limitations}

Dense packed evaluation is limited by the range of its exponent-field
encoding. At effective encoded base-two exponents of $-127$ and $-126.5$,
the packed instruction sequence produces NaN. Deployment therefore requires enforcing the range
contract or using online native-exponential evaluation. Our benchmark
checks use one seeded Gaussian fixture per length. Extending these observations to
general output and gradient guarantees requires bounds on approximation,
accumulation, and rounding over the intended input distribution.

Packed integer reduction is order-independent while its components
stay within range. However, the pass also includes floating-point
reductions and recovery paths, so full-pass determinism remains to
be evaluated.

We measure the combined implementation with prepared TLX launches and
FA4's CuTe launch path. Provider-specific launch costs contribute to
the timings. We have yet to ablate individual transformations in the benchmarked
binaries, isolate the cost of triggered recovery, compare compilation times,
or evaluate a CuTe implementation.

\section{Evaluation}
\label{sec:evaluation}

\subsection{FA4 benchmark configuration}
\label{sec:fa4-config}

We evaluate our implementation on a B200 at 1,000 W with the
variable-batch grid and original timer from FA4's
\code{benchmark\_attn.py}. Every sequence in a case has the same length;
batch size varies between cases. Inputs are BF16 with $d=128$, $H=16$,
$BN=32{,}768$, and sequence lengths 1,024, 2,048, 4,096, 8,192, 16,384,
and 32,768. Dense and causal forward and backward give 24 cases.
Gaussian inputs use seed zero, with the same fixture reused across
masks, providers, and replays at each length. Inputs use $[B,N,H,d]$
storage without layout-conversion copies. Forward runs in training
mode, including the state saved for backward.

We use PyTorch~\cite{paszke2019pytorch} \code{2.10.0+cu130} and the Triton 3.7.3 host runtime
to execute TLX binaries compiled with Triton \code{3.8.0+fb}.
TLX launches use prepared descriptors and argument bindings;
FA4 uses its CuTe launch path. Backward timing includes preprocessing,
gradient initialization, the main kernel, and required postprocessing.
Forward includes recovery where the guarded route uses it.

Our FA4 reference uses source revision \code{62892fe4},
CuTe DSL~\cite{nvidiacutlass} 4.6.2, and Quack 0.6.4; its compiler build reports CUDA 13.3.
We apply short-prefix backward corrections and a dense row-sum reordering.
These reference patches are included in the timed calls.
FA4 and TLX both pass all 24 dimension-128 cases.

FA4's paper reports CuTe DSL 4.4.1 and CUDA 13.1.
Its body identifies B200, while Appendix A.1 names B100; our device
records identify B200 for every measurement here. The exact source
revision used for the published measurements is unspecified.

The benchmark counts $4BHN^2d$ FLOPs for dense forward and $10BHN^2d$
for dense backward, including score recomputation; causal mode uses
half of each count. Throughput is this count divided by latency.
Each reported latency is the median of six mean latencies returned by
the original timer, one for each provider-order permutation.
Two identical FA4 callbacks provide a repeat control;
we use the same designated repeat as the reference in every case.
Aggregate speedups are geometric means of per-case
FA4-latency/TLX-latency ratios.

\subsection{FA4 variable-batch grid}

\begin{table}[htbp]
\centering
\caption{Per-length throughput on the dimension-128 FA4 variable-batch
grid at 1,000 W, current FA4 source/build. All entries use the original
timer and are in TFLOP/s. Each row contains a forward and a backward
pair, covering all 24 cases.}
\label{tab:fa4-per-length}
\small
\begin{tabular}{rrlrrrr}
\toprule
 & & & \multicolumn{2}{c}{Forward} & \multicolumn{2}{c}{Backward} \\
$N$ & $B$ & mask & FA4 & TLX & FA4 & TLX \\
\midrule
1,024 & 32 & dense & 1,373.7 & 1,558.2 & 822.9 & 982.2 \\
1,024 & 32 & causal & 719.6 & 692.4 & 485.3 & 555.3 \\
2,048 & 16 & dense & 1,467.0 & 1,659.0 & 1,061.1 & 1,223.1 \\
2,048 & 16 & causal & 1,019.7 & 995.9 & 750.8 & 850.3 \\
4,096 & 8 & dense & 1,520.5 & 1,715.0 & 1,227.8 & 1,363.1 \\
4,096 & 8 & causal & 1,244.0 & 1,230.6 & 985.3 & 1,081.5 \\
8,192 & 4 & dense & 1,548.9 & 1,748.4 & 1,330.1 & 1,453.7 \\
8,192 & 4 & causal & 1,381.9 & 1,387.4 & 1,161.1 & 1,261.6 \\
16,384 & 2 & dense & 1,566.2 & 1,770.6 & 1,391.9 & 1,425.7 \\
16,384 & 2 & causal & 1,466.6 & 1,452.8 & 1,282.4 & 1,355.6 \\
32,768 & 1 & dense & 1,582.5 & 1,786.8 & 1,249.7 & 1,326.7 \\
32,768 & 1 & causal & 1,510.2 & 1,488.9 & 1,355.7 & 1,317.8 \\
\bottomrule
\end{tabular}
\end{table}

\begin{table}[htbp]
\centering
\caption{TLX Attention on the FA4 variable-batch grid, B200 at
1,000 W, CuTe DSL 4.6.2, original benchmark timer. TFLOP/s values are geometric
means over six lengths for each pass and over all 24 cases in the final row.
All 24 cases pass numerical checks.}
\label{tab:fa4-latest}
\small
\begin{tabular}{lrrr}
\toprule
pass & FA4 & TLX Attention & TLX gain \\
\midrule
dense forward & 1,508.1 & 1,704.5 & 13.0\% \\
dense backward & 1,163.5 & 1,285.1 & 10.4\% \\
causal forward & 1,186.8 & 1,168.6 & $-1.5\%$ \\
causal backward & 947.7 & 1,023.7 & 8.0\% \\
\midrule
all 24 cases & 1,185.3 & 1,272.3 & 7.3\% \\
\bottomrule
\end{tabular}
\end{table}

Dense-forward gains range from 12.8 to 13.4 percent across all six
lengths (Table~\ref{tab:fa4-per-length}), so the gain
persists at long sequences. To relate this comparison to the published
baseline, we also examine FA4's peak: our reference reaches
1,582.5 TFLOP/s, 1.9 percent below the paper's 1,613 TFLOP/s.
This peak occurs at 32,768 tokens, where TLX reaches 1,786.8 TFLOP/s.

Causal attention benefits less. Combining directions gives gains of
11.7 percent for dense attention and 3.1 percent for causal attention,
yielding 7.3 percent overall (Table~\ref{tab:fa4-latest}), but causal
forward remains 1.5 percent slower. Its shorter tile loops leave less
independent tensor work to hide weight formation. PV therefore continues
to wait for weight formation after rescaling is removed.

Table~\ref{tab:gb300-summary} summarizes the GB300 measurements.

\begin{table}[htbp]
\centering
\small
\setlength{\tabcolsep}{5pt}
\begin{tabular}{lrrrr}
\toprule
Mask & FWD & BWD & Overall standalone & Joint FWD+BWD \\
\midrule
Dense & $0.990100\times$ & $1.102829\times$ & $1.044946\times$ & $1.073767\times$ \\
Causal & $1.012363\times$ & $1.101785\times$ & $1.056128\times$ & $1.078758\times$ \\
Both & $1.001170\times$ & $1.102307\times$ & $1.050522\times$ & $1.076260\times$ \\
\bottomrule
\end{tabular}
\caption{GB300 BF16, $d=128$, MHA 16/GQA 16/2, 8K/16K, complete round8 cohort:
paired geometric-mean FA4/TLX time ratios ($>1$ favors TLX).
Backward-derived ratios are diagnostic: some gradient checks fail, and only
TLX's GQA gradient reduction is excluded from timing.}
\label{tab:gb300-summary}
\end{table}

\subsection{TritonBench}

The archived TritonBench~\cite{tritonbench} experiment compares a different dense workload
against several implementations. It uses \code{blackwell\_attentions}
on B200 with a 750 W power cap and BF16 inputs, $B=4$, $H_q=H_{kv}=48$, $d=128$,
and $[B,H,N,d]$ storage over lengths 1,024, 2,048, 4,096, 8,192, and
16,384. Its benchmark-native CuTeDSL backend and public FA4 are separate
comparators.
The suite also includes cuDNN~\cite{chetlur2014cudnn}.

The denoising wrapper requests a 1,965 MHz graphics-clock lock alongside
the power cap. Retained launch logs record this request, but the final
forward-reference report describes unlocked clocks; its execution clock
state is unresolved in the archive.

For the August 2026 dense configuration in Table~\ref{tab:tritonbench},
we estimate TLX throughput from profiler anchors and timing ratios
over 21 paired rounds, whereas comparator values come directly from
profiler measurements. The anchors use \code{-{}-rep 3000 -{}-sleep 1.0};
public FA4 was measured on a separate B200.

\begin{table}[htbp]
\centering
\caption{Recorded dense BF16 TritonBench results with a 750 W power cap. TFLOP/s values
are geometric means over five sequence lengths. TLX entries are
ratio-adjusted estimates for the August 2026 dense configuration.}
\label{tab:tritonbench}
\small
\begin{tabular}{lrrrrr}
\toprule
pass & TLX Attention & CuTeDSL & cuDNN & public FA4 & TLX vs CuTeDSL \\
\midrule
forward & 1,127.7 & 994.4 & 960.4 & 987.9 & 13.4\% \\
backward & 998.8 & 803.0 & 732.3 & 757.7 & 24.4\% \\
\bottomrule
\end{tabular}
\end{table}

For this dense workload, the minimum estimated geometric-mean gains
over the other evaluated implementations are
13.4 percent in forward and 24.4 percent in backward.

\section{Related Work}

FlashAttention kernels avoid materializing score and probability matrices in
HBM through tiling~\cite{dao2022flashattention,dao2023flashattention2,shah2024flashattention3,dao2026flashattention4}.
Online softmax supplies stable streaming normalization
~\cite{milakov2018online}. Chunked attention with checkpointed recomputation
also avoids materializing the quadratic score matrix during differentiation
~\cite{rabe2021selfattention}. I/O analyses establish cache-regime-dependent
communication bounds for exact forward and backward
attention~\cite{saha2024io,li2024backwardio}.

FlashDecoding++ uses a fixed, precomputed ``unified max value'' to
remove synchronized rescaling from decoding~\cite{hong2023flashdecoding}.
Our use of a fixed shift builds on that idea; the differences here are
training forward and backward, packed BF16 exponential evaluation, and
reuse of the inverse denominator in backward. VFA reaches a fixed shift
through a different update schedule: it initializes the maximum from
key summaries, updates it on a subset of blocks, and then freezes
it~\cite{sun2026vfa}. These approaches change when normalization state
is updated, while exponent-field approximation changes the arithmetic
used to generate weights.

Schraudolph introduced exponent-field approximations to the
exponential~\cite{schraudolph1999}. VEXP applies related BF16
constructions to softmax and FlashAttention-2 on a RISC-V floating-point
cluster~\cite{wang2025vexp}, whereas Softermax combines base-two,
low-precision softmax with online normalization in a hardware/software
co-design~\cite{stevens2021softermax}.

SageAttention3 studies microscaling FP4 attention for inference and explores
8-bit training~\cite{zhang2025sageattention3}. Qiu and Yao identify
a source of error relevant to our prefix correction:
rounding the saved output can make
$D=\operatorname{rowsum}(O\odot dO)$ differ from reduction through
reconstructed probabilities~\cite{qiu2026lowprecision}. This distinction
underlies the alternative $D$ computation in our short causal prefixes.

Persistent and work-centric schedules reduce tile quantization loss in
GEMM and long-context decode attention~\cite{osama2023streamk,sanovar2024lean}.
FlashInfer plans persistent attention work from sequence lengths and
orders tiles by descending cost~\cite{ye2025flashinfer}. DASH studies
attention-backward scheduling with deterministic $dQ$
reduction~\cite{qiang2026dash}; its ordered floating-point accumulation
differs from the range-valid packed integer reduction used here.

Colfax's dimension-64 backward analysis separates score and probability
storage so that scores can be released earlier~\cite{colfax2026backward}.
Its decode analysis instead alternates score/probability buffers to overlap
the next QK with the current softmax~\cite{colfax2026pingpong}.
Both schedules exploit spare \tmem, whereas our measured forward
retains the aliased score/probability storage in
Figure~\ref{fig:forward-overlap-dependencies}.

\section{Conclusion}

Fixing the dense row shift removes recurrent accumulator corrections,
and saving the inverse denominator moves backward normalization into
preprocessing. The combined implementation improves geometric-mean throughput
over the FA4 reference by 7.3 percent across all 24 numerically qualified
dimension-128 cases on B200 at 1,000 W, using the original timer and
prepared launches.

Causal forward remains 1.5 percent slower.
Because PV still waits for probability publication, removing rescaling
leaves a dependency exposed on short causal tiles. Overlapping more
probability work with independent tensor operations is therefore a
remaining scheduling opportunity within the available registers and
on-chip storage.

\paragraph{Artifact availability.}
Source code, binary manifests, benchmark records, and numerical-check
scripts are retained in an internal archive. A public artifact release
has not been prepared for this draft.

These algorithmic and scheduling ideas are transferable to other GPU
programming languages and frameworks. The results compare specific
implementations and do not establish Triton or TLX as inherently better
languages. Future work will integrate these ideas into the CuTe implementation
of FlashAttention-4.

\end{document}